\documentclass{article}

\usepackage{microtype}
\usepackage{graphicx}
\usepackage{subfig}
\usepackage[T1]{fontenc}
\usepackage{booktabs}
\usepackage{hyperref}

\usepackage[accepted]{icml2024}

\usepackage{amsmath}
\usepackage{amssymb}
\usepackage{mathtools}
\usepackage{amsthm}
\usepackage{dsfont}

\usepackage[shortlabels]{enumitem}
\setlist[enumerate]{nosep}
\setlist[itemize]{nosep}

\usepackage{setspace}
\AtBeginDocument{
  \setlength{\abovedisplayskip}{3pt plus 2pt minus 2pt}
  \setlength{\belowdisplayskip}{3pt plus 2pt minus 2pt}
}

\usepackage{float}
\usepackage{caption}
\usepackage[capitalize,noabbrev]{cleveref}

\theoremstyle{plain}
\newtheorem{theorem}{Theorem}[section]
\newtheorem{proposition}[theorem]{Proposition}

\theoremstyle{definition}

\theoremstyle{remark}

\usepackage[textsize=tiny]{todonotes}

\icmltitlerunning{ELA: Exploited Level Augmentation for Offline Learning in Zero-Sum Games}

\begin{document}

\twocolumn[
\icmltitle{ELA: Exploited Level Augmentation for Offline Learning in Zero-Sum Games}

\icmlsetsymbol{equal}{*}

\begin{icmlauthorlist}
\icmlauthor{Shiqi Lei}{equal,casia}
\icmlauthor{Kanghoon Lee}{equal,kaist}
\icmlauthor{Linjing Li}{casia}
\icmlauthor{Jinkyoo Park}{kaist}
\icmlauthor{Jiachen Li}{ucr}
\end{icmlauthorlist}

\icmlaffiliation{casia}{Institute of Automation, Chinese Academy of Sciences, Beijing, China}
\icmlaffiliation{kaist}{Department of Industrial \& Systems Engineering, KAIST, Daejeon, South Korea}
\icmlaffiliation{ucr}{Department of Electrical and Computer Engineering, University of California, Riverside, California, USA}

\icmlcorrespondingauthor{Jiachen Li}{jiachen.li@ucr.edu}

\icmlkeywords{offline learning, imitation learning, offline reinforcement learning, representation learning, zero-sum games}

\vskip 0.3in
]

\printAffiliationsAndNotice{\icmlEqualContribution}

\begin{abstract}
Offline learning has become widely used due to its ability to derive effective policies from offline datasets gathered by expert demonstrators without interacting with the environment directly. Recent research has explored various ways to enhance offline learning efficiency by considering the characteristics (e.g., expertise level or multiple demonstrators) of the dataset. However, a different approach is necessary in the context of zero-sum games, where outcomes vary significantly based on the strategy of the opponent. In this study, we introduce a novel approach that uses unsupervised learning techniques to estimate the exploited level of each trajectory from the offline dataset of zero-sum games made by diverse demonstrators. Subsequently, we incorporate the estimated exploited level into the offline learning to maximize the influence of the dominant strategy. Our method enables interpretable exploited level estimation in multiple zero-sum games and effectively identifies dominant strategy data. Also, our exploited level augmented offline learning significantly enhances the original offline learning algorithms including imitation learning and offline reinforcement learning for zero-sum games.
\end{abstract}

\section{Introduction}

Although reinforcement learning has become a powerful technique extensively applied for decision making in various domains such as robotic manipulation, autonomous driving, and game playing \cite{andrychowicz2020learning, chen2019model, vinyals2019grandmaster}, conventional reinforcement learning demands substantial online interactions with the environment, which is a process that can be both costly and sample inefficient while potentially leading to safety risks \cite{berner2019dota, bojarski2016end}.
To address these issues, many methods have emerged to enable efficient learning using offline datasets generated by demonstrators. For example, behavior cloning \cite{pomerleau1988alvinn} replicates actions from the offline dataset, assuming demonstrators are experts. It offers ease of implementation and learning but is sensitive to suboptimal demonstrations and limited generalization. 
In contrast, offline reinforcement learning \cite{fujimoto2019off, kumar2020conservative} aims to derive an optimal policy from the dataset. While it exhibits robustness to suboptimal data and generalization issues, it poses challenges when dealing with small or biased datasets.

\begin{figure}[t]
  \centering
  \includegraphics[width=1.\linewidth]{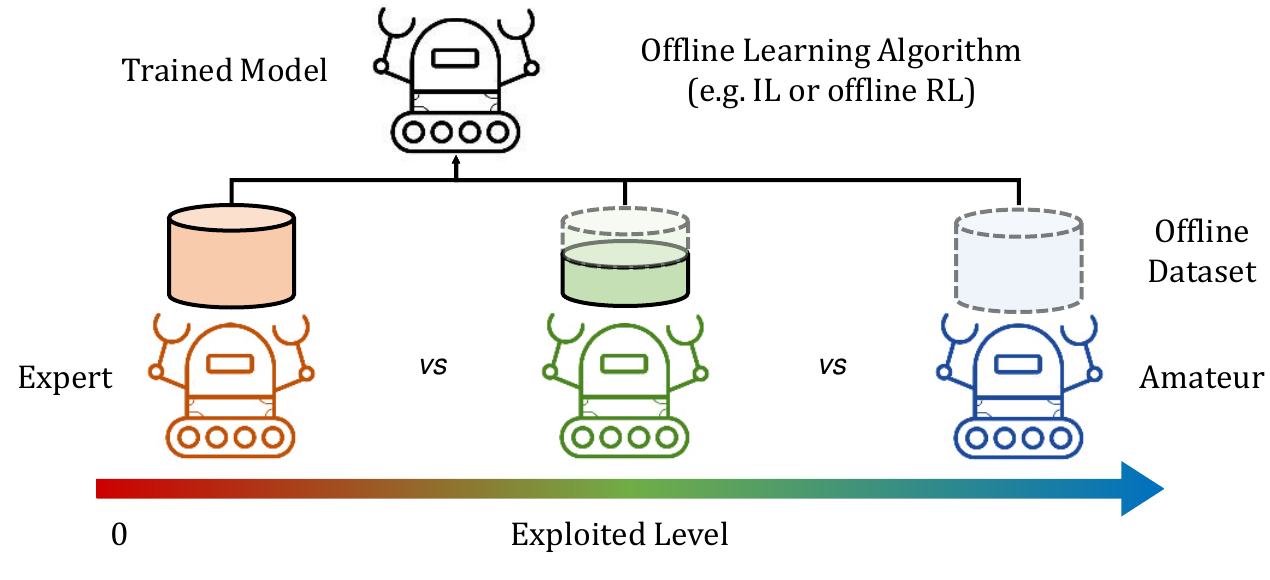}
  \vspace{-0.65cm}
  \caption{Illustration of ELA (Exploited Level Augmentation) for offline learning in zero-sum games. It learns the exploited level in an unsupervised manner and prioritizes trajectories created by the dominant behavior in offline learning.}
  \label{fig:intro}
\end{figure}

In real-world scenarios, offline datasets exhibit diversity in various aspects. One notable aspect concerns the coverage of the labels (state, action, and reward) in the data. Ideally, data should contain full pairs of states, actions, and rewards for effective offline learning. However, practical datasets often comprise sequences of state-action pairs without reward labels or, in some cases, only states \cite{dai2023learning, ettinger2021large}. 
To address these challenges, several methods have been proposed, including techniques for labeling via inverse dynamic model \cite{baker2022video, lu2022imitation} or reward labeling \cite{yu2022leverage}.
Another aspect of diversity is related to the characteristics of the demonstrators for generating the dataset. For example, demonstrators can vary in terms of task execution methods or skill levels. More specifically, different demonstrators may exhibit distinct preferences and employ various approaches even when performing the same task, resulting in multi-modal datasets \cite{shafiullah2022behavior}.
Meanwhile, diversity may arise from different skill levels among demonstrators, which not only introduces multi-modal challenges but also complicates the identification of irrelevant data that hinders the learning process \cite{mandlekar2021matters}.
To mitigate these issues, \citet{pearce2023imitating} includes models for handling demonstrator multi-modality and \citet{beliaev2022imitation} proposes a method to distinguish highly skilled demonstrators.

While there has been significant research on learning from offline data in multi-agent systems \cite{pan2022plan}, handling the unique characteristics of the demonstrators is still a largely underexplored area.
Specifically, in competitive environments such as zero-sum games, the data distribution is influenced by the attributes and expertise of the participating players. 
Therefore, unlike single-agent offline data, it is crucial to extract a suitable representation by considering the individual characteristics of each participant in the game.
Additionally, because the strategies in such games can take diverse forms \cite{czarnecki2020real}, successful learning from offline data in zero-sum games requires a deep understanding and consideration of these factors.

In this work, we introduce ELA (i.e., Exploited Level Augmentation) for offline learning designed to learn from low-exploited behavioral data. ELA discerns exploited levels within diverse demonstrators' offline datasets in zero-sum games. Figure \ref{fig:intro} illustrates an overview of ELA.

The main contributions of this paper are as follows:
\begin{itemize}
\item We propose the Partially-trainable-conditioned Variational Recurrent Neural Network (P-VRNN) and an unsupervised framework for learning strategy representation of trajectories in multi-agent games.
\item We define the EL (i.e., Exploited Level) of the strategies, and propose an unsupervised method for estimating the exploited level in the offline dataset generated by various demonstrators for zero-sum games. 
\item We introduce ELA, a technique for offline learning that incorporates the exploited level of each trajectory. It is compatible with various offline learning algorithms.
\item We demonstrate that our EL estimator serves as an effective indicator in zero-sum games, including Rock-Paper-Scissors (RPS), Two-player Pong, and Limit Texas Hold'em. ELA significantly enhances both imitation and offline reinforcement learning performance.
\end{itemize}

\section{Related Work}

In the early days of learning from demonstrations (LfD) or imitation learning (IL), behavior cloning (BC) \cite{bain1995framework} and inverse reinforcement learning (IRL) \cite{russell1998learning} were proposed to learn behavior policies from offline data. 
In recent years, lots of different methods have emerged, such as DAgger \cite{ross2011reduction}, GAIL \cite{ho2016generative}, and other variants based on BC or IRL \cite{ziebart2008maximum}. 
However, some of them (e.g., DAgger) require the experts to make real-time decisions when encountering new situations rather than solely relying on offline data.
Some other methods (e.g., GAIL) require knowledge of the environment dynamics, and many methods follow its structure \cite{ding2019goal}.
Offline reinforcement learning \cite{ernst2005tree} emerged and raised with the development of reinforcement learning, and many methods have been proposed in recent years \cite{fujimoto2019off,kumar2020conservative}.
Nevertheless, offline RL demands a reward for each time step, imposing significant constraints on its applicability. 
Since our augmentation method does not have such requirements, we only discuss methods that use the trajectory information and terminal reward to deal with suboptimal demonstrations in the following.

\textbf{IRL-based methods.} \citet{brown2019extrapolating} proposed an IRL-based Trajectory-ranked Reward Extrapolation (T-REX) algorithm, which extrapolates approximately ranked demonstrations, so that better-designed reward functions can be derived from poor demonstrations.
After that, Distrubance-based Reward Extrapolation (D-REX) \cite{brown2020better} generates ranked demonstrations by introducing noise into a BC-learned policy and leveraging T-REX.
\citet{chen2021learning} highlighted a limitation of D-REX that Luce's rule inaccurately depicts the noise-performance relationship and proposed a method to minimize the effect of suboptimal demonstrations by generating optimality-parameterized data.
While IRL-based methods outperform demonstrations with limited expertise, they face challenges in zero-sum games, where agent interactions with the environment and opponents significantly influence the terminal reward.

\textbf{BC-based methods.}
\citet{sasaki2020behavioral} enhances BC for noisy demonstrations, while TRAIL \cite{yang2021trail} achieves sample-efficient imitation learning via a factored transition model.
Play-LMP \cite{lynch2020learning} leverages unsupervised representation learning in a latent plan space for improved task generalization. However, employing a variational auto-encoder (VAE) with the encoder outputting latent plans is unsuitable for zero-sum games, potentially leaking opponent information from the observations and disrupting the evaluation of the demonstrator.

\textbf{IL with representation learning.} 
The work by \citet{beliaev2022imitation} closely aligns with our research, sharing the primary goal of extracting expertise levels of trajectories.
They assumed that the demonstrator has a vector indicating the expertise of latent skills, with each skill requiring a different level at a specific state. These elements jointly derive the expertise level.
The method also considers the policy worse when it is closer to uniformly random distribution.
However, this assumption cannot be satisfied even in simple games like Rock-Paper-Scissors (RPS), where a uniformly random strategy constitutes a Nash equilibrium.
\citet{grover2018learning} also studied learning policy representations, but they used the information of agent identification during training, which enables them to add a loss to distinguish one agent from others.
In our work, we propose a method for policy representations that effectively captures optimality in a zero-sum game without relying on agent identification.

\section{Preliminaries}\label{sec:preliminaries}
Components of an $N$-player zero-sum game are as follows:
\begin{itemize}
    \item Player set $P$: $P=\{1,...,N\}$; 
    \item State $s$: all the information at a certain status, including action history and imperfect information;
    \item Observation $o_i(s)$: all the information player $i$ can get at a certain state $s\in S$;
    \item Action space $A_i(o_i)$: all actions that can be done at a certain observation $o_i$;
    \item Terminal states $z\in Z\subset S$: all states that no further actions can be done;
    \item Rewards $r_i: Z \to \mathbb{R}$: the reward given to player $i$, and $\sum_{i\in P}r_i(z)=0$, $\forall z\in Z$;
    \item Strategy $\pi_i(a \mid o_i)$: the probability of player $i$ choosing action $a$ at observation $o_i$.
\end{itemize}

$\Pi_i$ is the set of all strategies of player $i$. In symmetric cases, we use $\Pi$ to denote the strategy set, i.e. $\Pi_i=\Pi$, $\forall i$.
We define the reach probability
$$p_{\pi}(s)=\prod_{(s',a')\to s}\pi(a'\mid o_i(s'))$$
as the probability of reaching state $s$ with strategy $\pi$, where $(s',a')\to s$ means that choosing action $a'$ at $s'$ is the choice of going to state $s$. 
Then, we can naturally define the expected reward of player $i$ with strategy $\pi$ as
\begin{align*}
    r_i(\pi)=\sum_{z\in Z}p_{\pi}(z)r_i(z).
\end{align*}
We use $r_i(\pi_{-i},\pi_i)$ to specify the player strategy $\pi_i$ and opponent strategy $\pi_{-i}$. The best response of opponent strategy $\pi_{-i}$ is defined as $BR(\pi_{-i})=\text{argmax}_{\pi_i'}r_i(\pi_{-i}, \pi_i')$. 
We additionally define the best response of strategy $\pi_i$ as 
\begin{align*}
    BR(\pi_i)=\text{argmax}_{\pi_{-i}'}\sum_{j\in P,j\neq i}r_j(\pi_{-i}',\pi_i),
\end{align*}
which equals to $\text{argmin}_{\pi_{-i}'}r_i(\pi_{-i}',\pi_i)$ in zero-sum case. 
We define the exploitability of strategy $\pi$ as 
\begin{align*}
    E(\pi)=\sum_{i\in P}\left(r_i(\pi_{-i},BR(\pi_{-i}))-r_i(\pi)\right).
\end{align*} 
In zero-sum symmetric cases, we define the exploitability of a player strategy $\pi_i$ as 
\begin{align*}
    E(\pi_i)=-r_i(BR(\pi_{i}),\pi_i) =\sum_{j\in P,j\neq i}r_j(BR(\pi_{i}), \pi_i).
\end{align*}

\section{An Intuition of Exploited Level}

\begin{figure}[t]
  \centering
  \includegraphics[width=0.9\columnwidth]{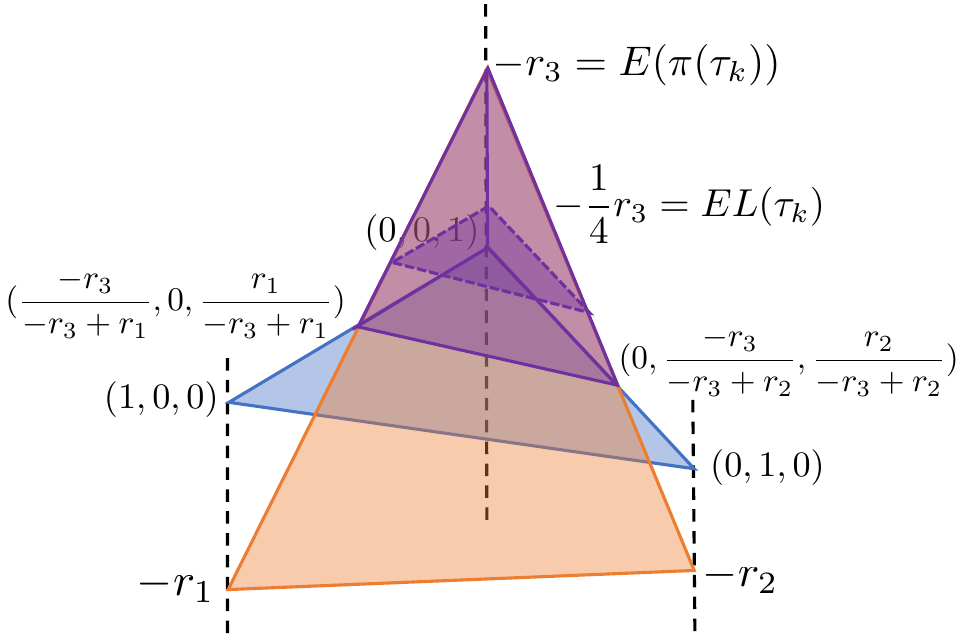}
  \vspace{-0.3cm}
  \caption{Illustration of EL and exploitability of a strategy in a two-player zero-sum game with three pure strategies.}
  \vspace{0.3cm}
  \label{fig:toy}
\end{figure}

In this section, we provide an intuition of \textit{Exploited Level} (EL) with a toy model. It serves as a proportional approximation of exploitability with a certain distribution on the strategy set. 
Consider a 2-player zero-sum symmetric game that has $n$ pure strategies $\pi_i$, $i=1,...,n$. 
All strategies are convex combinations of pure strategies, i.e., $\pi =\sum_{i=1}^{n}a_i \pi_i$, where $\sum_{i=1}^{n}a_i=1$, $0\leq a_i \leq 1, \forall i$.
For simplicity, we assume that each trajectory $\tau$ can be directly mapped to a strategy $\pi(\tau)$. 
In our setting, where all the players are competent, each player can be exploited by at most one pure strategy. 
As for the overall strategy distribution over $\Pi$, we assume the $(a_1, a_2, ..., a_n)$ has uniform distribution over $(n-1)$-dimensional standard simplex.

The definition of EL is as follows:
$$EL(\tau)=\mathbb{E}_{\pi}\left[-r(\pi,\pi(\tau))\mid r(\pi,\pi(\tau))\leq 0\right].$$
For a trajectory $\tau_k$, let $r(\pi_i,\pi(\tau_k))=r_i$. By our assumption, only one $j\in\{1,2,...,n\}$ satisfies that $r_j<0$, while $r_i\geq 0, \forall i\neq j$. We can directly see that $E(\pi(\tau_k))=-r_j$. 

Since $EL(\tau_k)$ is a conditional expectation defined on $\Pi$, we can view it as a conditional expected value over an $(n-1)$-dimensional simplex. 
When $\pi =\sum_{i=1}^{n}a_i \pi_i$, $r(\pi,\pi(\tau_k))=\sum_{i=1}^{n}a_i r_i$, the condition becomes $\sum_{i=1}^{n}r_i a_i\leq 0$. 
Thus, the expectation is still defined over an $(n-1)$-dimensional simplex, but a smaller one, with vertices $\{(0,...,a_i=\frac{-r_j}{-r_j+r_i},...,a_j=\frac{r_i}{-r_j+r_i},...,0)\mid \forall i\neq j\}\cup \{(0,...,a_j=1,...,0)\}$. 
Then, we can consider adding another dimension on the simplex, so that the new dimension has value $-r(\pi,\pi(\tau))$. Due to linearity, the new object becomes an $n$-dimensional pyramid, and the desired expectation is the height of the pyramid's centroid w.r.t. the surface of the original $(n-1)$-dimensional simplex. 
From calculus, the height of the centroid of $n$-dimensional pyramid is always $\frac{1}{n+1}$ of the height of the pyramid w.r.t. its base. Since the height is $r_j$, the expectation is $\frac{1}{n+1}r_j$. 
So $$EL(\tau_k)=\frac{1}{n+1}E(\pi(\tau_k))$$ always holds in this case, which shows that EL is an appropriate indicator. A strategy of a game with three different pure strategies is shown as an example in Figure \ref{fig:toy}, with EL and exploitability visualized.

Concretely, consider an RPS game (see details in Appendix \ref{app:games}) and let $\pi_1,\pi_2$ and $\pi_3$ be the pure strategies of choosing rock, paper, and scissors, respectively. Let the strategy of $\tau$ be $\pi(\tau)=(0,2/3,1/3)$, i.e. "choosing paper with $2/3$ probability and choosing scissors with $1/3$ probability". Then we can easily derive that $-r_1=-r_2=-1/3, -r_3=2/3$. So we have $E(\pi(\tau))=2/3$, while $EL(\tau)=1/6$.

\section{Problem Formulation}

Consider a zero-sum game and we have a dataset of game histories, including the trajectories of each player and terminal rewards. The trajectories are generated by diverse players, ranging from high-level experts to amateurs. We aim to distinguish the players with different levels and learn an expert policy from the dataset via offline learning. We assume that we do not have the demonstrator identification.

\begin{figure}[t]
  \centering
  \includegraphics[width=0.9\columnwidth]{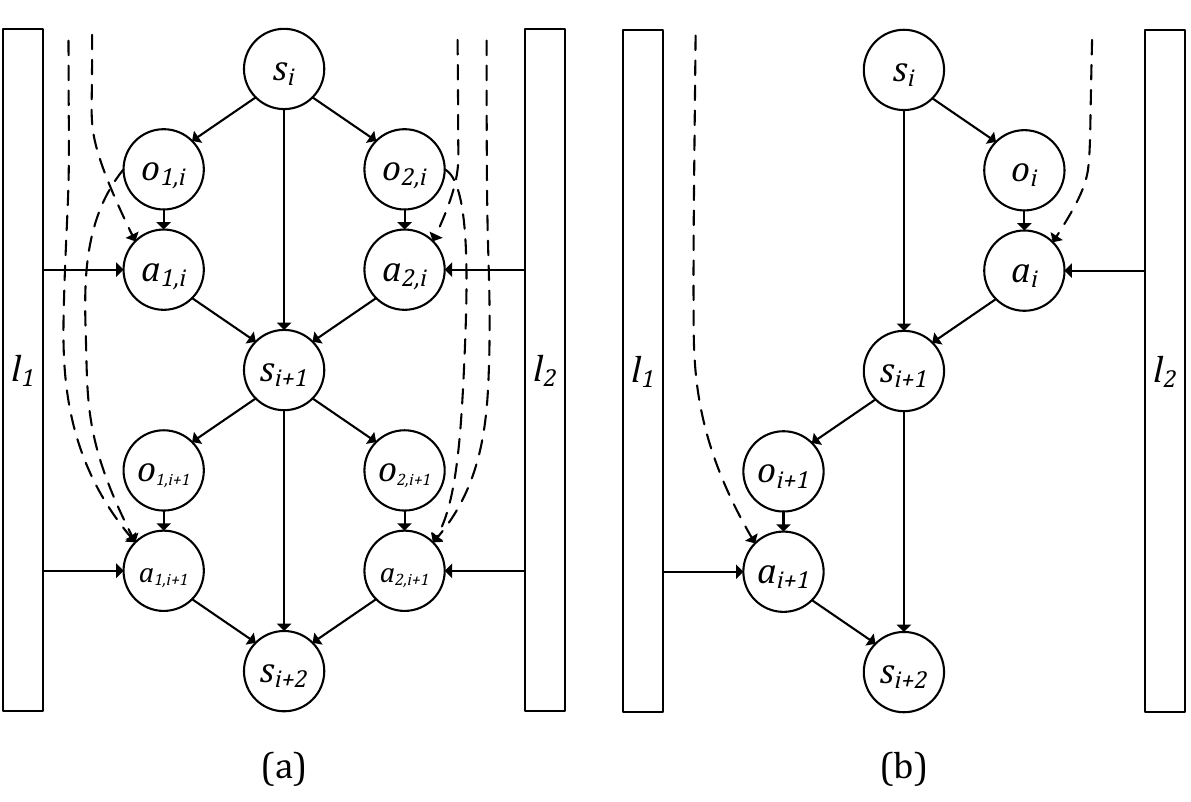}
  \vspace{-0.3cm}
  \caption{The basic structures of games with representation-dependent policy. (a) Games with simultaneous actions. (b) Games with sequential actions.}
  \vspace{0.3cm}
  \label{fig:basic}
\end{figure}

In our problem, the trajectories
\begin{align*}
    \tau^{i,j}=((o_0^{i,j},a_0^{i,j}),...,(o_{T^{i,j}}^{i,j},a_{T^{i,j}}^{i,j}))
\end{align*}
are collected for each player $i$ and game $j$, in which $o_t^{i,j}$ is the observation, $a_t^{i,j}$ is the action at time $t$ and $T^{i,j}$ is the length of the corresponding trajectory. 
The dataset of trajectories of $M$ games is denoted as
\begin{align*}
    \Gamma=\{\tau^{i,j}\mid i=1,...,N,j=1,...,M\}.
\end{align*}
We remark that the observation of each player is different in an imperfect information game. The terminal reward $r^{i,j}$ is recorded for each trajectory. 
For simplicity, we later omit superscripts $i,j$ when referring to a single trajectory. We assume that within a single trajectory, the strategy of a player is consistent. 
The probability of a certain action following a certain sequence of observations is denoted as $\pi(a\mid l, o_0,...,o_t)$, where $l$ represents the strategy representation vector of the trajectory. The basic structures of games with strategy representation are illustrated in Figure~\ref{fig:basic}.

\section{Method}
\begin{figure*}[t]
    \centering
    \includegraphics[width=\textwidth]{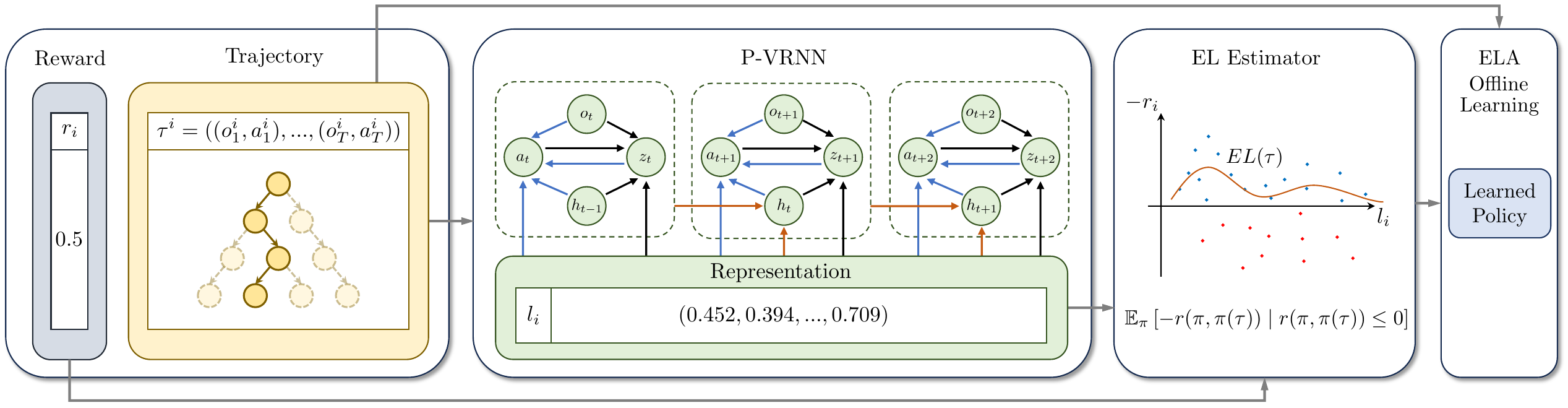}
    \vspace{-0.7cm}
    \caption{The overall diagram of Exploited Level Augmentation (ELA) for offline learning.}
    \vspace{-0.3cm}
    \label{fig:overview}
\end{figure*}

Evaluating a player's strategy from a specific trajectory in a dataset of trajectories and terminal rewards of a game is challenging. 
However, estimating individual strategies becomes feasible with unique player identification in the dataset, consequently allowing for the evaluation of each player's skill level under the assumption of a consistent strategy. 
Nevertheless, given the absence of player identification in our setting, discerning the strategy directly from trajectories is not possible.
To address this limitation, given a prior distribution of strategies, we can derive the probability distribution of a trajectory's strategy.
Consequently, acquiring representations of trajectories that illustrate a distribution over the strategy space $\Pi$ emerges as a feasible solution.
After obtaining the strategy representation, the terminal reward can be utilized to estimate how well the player does using the proposed estimator. These estimations, in turn, can be leveraged to enhance offline learning algorithms by prioritizing relevant data.
Our method consists of three major procedures, as illustrated in Figure \ref{fig:overview}:
\begin{enumerate}
\setlength{\itemsep}{0mm}
    \item Obtaining the strategy representation with unsupervised learning;
    \item Deriving the function of obtaining the exploited level from the strategy representation;
    \item Exploited level augmented offline learning.
\end{enumerate}

\begin{figure}[!]
  \centering
  \includegraphics[width=0.9\linewidth]{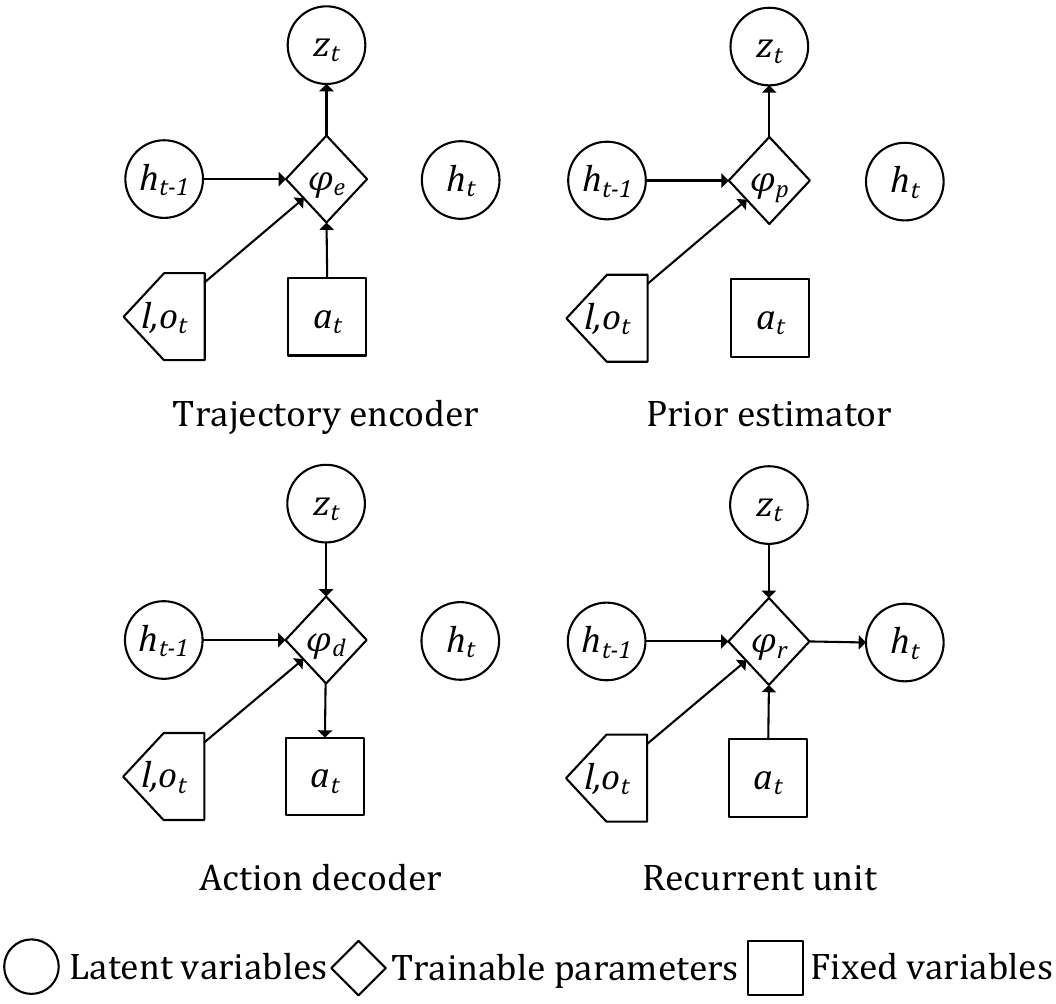}
  \vspace{-0.3cm}
  \caption{The network structure of the P-VRNN model.}
  \label{fig:network}
\end{figure}

\subsection{Learning Strategy Representation}

We propose a Partially-trainable-conditioned Variational Recurrent Neural Network (P-VRNN) where partially-trainable-conditioned means that part of the condition on the neural network is trainable, and the condition acquisition process is entirely unsupervised. 
The proposed P-VRNN has four major components similar to the VRNN with an additional condition, as shown in Figure \ref{fig:network}.
Instead of using the latent variable as an indicator \cite{lynch2020learning}, we use part of the condition as a representation vector. 

\textbf{Trajectory encoder.} The trajectory encoder obtains the latent variable $z_t$ at time $t$ from the past trajectory information including action $a_t$. The encoder uses the current action $a_t$, strategy representation $l$, and all the past observations $\{o_i\}_{i=1}^t$ according to Figure \ref{fig:basic}. 
It can be seen from the computation graph of P-VRNN that the last step recurrent variable $h_{t-1}$ contains information of $\{o_i\}_{i=1}^{t-1}$, $\{a_i\}_{i=1}^{t-1}$ and $l$ since $z_t$ has the information of $(h_{t-1}, o_t, l)$ and $h_t$ gathers the information of $(h_{t-1}, o_t, a_t, z_t, l)$.
The information of $a_{i-1}$ is also contained in $o_i$, thus we define the trajectory encoder as
$$\phi_{\text{e}}(h_{t-1}, o_t, a_t, l)=[\mu_t^z, \sigma_t^z],$$
$$z_t\mid h_{t-1}, o_t, a_t, l \sim \mathcal{N}(\mu_t^z, \text{diag}((\sigma_t^z)^2)),$$
where $\phi_{\text{x}}$ is a trainable mapping and subscript $\text{x}$ can be any character. 
We follow the convention in VAE and assume that the latent variable has a diagonal covariance matrix.

\textbf{Prior estimator.} Without knowing the actual action $a_t$, the prior of latent variable $z_t$ can be derived only from the representation $l$ and the past observations $\{o_i\}_{i=1}^t$. 
We extract information from the past and obtain an estimate of the prior of $z_t$. 
We define the prior estimator as
$$\phi_{\text{p}}(h_{t-1}, o_t, l)=[\hat{\mu}_t, \hat{\sigma}_t],$$
$$z_t\mid h_{t-1}, o_t, l\sim \mathcal{N}(\hat{\mu}_t, \text{diag}(\hat{\sigma}_t^2)).$$

\textbf{Action decoder.} The action decoder works inversely and obtains action $a_t$ from the latent variable $z_t$, past observations $\{o_i\}_{i=1}^t$ and representation $l$, which can also be substituted by using $h_{t-1}$, $z_t$, $o_t$ and $l$. 
We obtain the prediction of actions by the action decoder, which is defined formally as
$$\phi_{\text{d}}(h_{t-1}, z_t, o_t, l)=[\mu_t^x, \sigma_t^x],$$
$$a_t\mid h_{t-1}, z_t, o_t, l \sim \mathcal{N}(\mu_t^x, \text{diag}((\sigma_t^x)^2)).$$

\textbf{Recurrent unit.} The recurrent unit takes in all the variables of the current step and the output of the recurrent unit of the last step, which extracts all the past information and passes it on to the next step. At each time step, the recurrent unit is updated by
$$\phi_{\text{r}}(h_{t-1}, a_t, z_t, o_t, l)=h_t.$$

We design the loss function of P-VRNN as follows:

\textbf{Reconstruction loss.} The encoder-decoder model aims to closely match the true data. In our dataset where each moment has only one one-hot encoded action, the reconstruction loss is defined as $\mathcal{L}_{\text{recon},t}=CE(\mu_t^z, a_t)$, and the variance $\sigma_t^z$ is omitted, where the cross entropy is expressed as $CE(p,q)=-\int p(x)\log q(x)\text{d}x$.

\textbf{Regularization loss.} The prediction of the prior estimator is expected to closely align with the result of the trajectory encoder. Since the output of both $\phi_{\text{p}}$ and $\phi_{\text{e}}$ are normal distributions, we follow the design of VAE that uses KL divergence. The regularization loss is thus defined as $\mathcal{L}_{\text{KL},t} = KL(\mathcal{N}(\mu_t^z, \text{diag}((\sigma_t^z)^2)) \parallel \mathcal{N}(\hat{\mu}_t, \text{diag}(\hat{\sigma}_t^2))$, which can be explicitly calculated by 
    $\mathcal{L}_{\text{KL},t}=\frac{1}{2}\left[\sum_{i=1}^k\left((\log\hat{\sigma}_t)_i-\log(\sigma_t^z)_i+(\hat{\sigma}_t)_i^{-1}w_i\right) - k\right]$,
    where $w_i=((\hat{\mu}_t)_i - (\mu_t^z)_i)^2+(\sigma_t^z)_i$ and $k$ is the dimension of the latent space.

Finally, the total loss is written as
$$\mathcal{L}=\sum_{t=1}^{T} \left(\mathcal{L}_{\text{recon},t}+\mathcal{L}_{\text{KL},t}\right).$$
When learning the strategy representation, the trainable random representation vector $l^{i,j}$ is initiated for each trajectory $\tau^{i,j}$.
The condition part of P-VRNN consists of observation $o_t$ which changes over time and the representation vector $l$ which is consistent during the whole trajectory and trainable.
During training, all the $l$'s are optimized together with the parameters of 
$\phi_{\text{p}}$, $\phi_{\text{e}}$, $\phi_{\text{d}}$, and $\phi_{\text{r}}$. 
The networks are trained to perform better in predicting the next step of trajectories, while the representations are optimized differently for each trajectory to provide customized predictions.
Consequently, the $l$ should be adjusted based on the tendency to express trajectory strategies more effectively. The entire process of obtaining $l$ is not only unsupervised but also without the information of players' identification.

\subsection{Exploited Level Estimator}\label{subsec:ele}

In two-player symmetric zero-sum games, it is common to use exploitability as a measure for evaluating the effectiveness of a strategy. 
However, it is extremely difficult to obtain exploitability with a single trajectory since we cannot: 1) infer or modify the strategy of the opponent; or 2) make any interaction with the environment.
For a strategy $\pi_i$, if we have many trajectories that have a strategy similar to it and the opponents use a large variety of strategies (so that there is one strategy near the best response), then $\forall \epsilon>0$, there exists a $\delta>0$ which satisfies the following approximation:
$$\left|E(\pi_i)-\max_{d(\pi_i',\pi_i)<\delta}\left[-r(\hat{\pi}_{-i},\pi_i')\right]\right|<\epsilon,$$
where $d$ is a distance over the strategy space.

However, if we have many trajectories so that for each trajectory, the opponent strategies can cover most kinds of strategies, and the trajectories with similar representation vectors have similar strategy distributions, can we still use the minimum reward of trajectories with representation near itself to serve as an approximation of negative exploitability? 
First, we define measure $\text{d}\pi$ on strategy space $\Pi$ according to the probability of $\pi$ chosen in the whole dataset:
$$\int_{\pi\in S}\text{d}\pi=\mathbb{P}\left[\tau \sim \pi, \pi \in S,\forall\tau\in\Gamma\right],$$
where $S$ is an arbitrary subset of $\Pi$.
Denote the trajectory as $\tau$, the representation function learned above as $f(\tau)$, and the reward of $\tau$ as $r(\tau)$. 
We remark that a trajectory $\tau$ should be mapped to a probability distribution of strategies such that $\int_{\pi\in\Pi}\tau(\pi)\text{d}\pi=1$, where $\tau(\pi)$ is the probability of using strategy $\pi$ when having trajectory $\tau$, instead of a single strategy. But we can view the mixture of $\pi$ with probability $\tau(\pi)$ as a single mixed strategy $\int_{\pi\in\Pi}\pi\tau(\pi)\text{d}\pi$, so we can still use notation $\pi(\tau)$ to represent the strategy of $\tau$. 
Using the above method, we can approximate $E(\pi(\tau))$, i.e.,
$$\left|E(\pi(\tau))-\max_{d(f(\tau'),f(\tau))<\delta}\left[-r(\tau')\right]\right|<\epsilon.$$
But the $E(\pi(\tau))$ we are approximating is \textbf{not} what we desire. 
In order to measure the exploitability of $\tau$, we should calculate $E(\tau):=\int_{\pi\in\Pi}\tau(\pi)E(\pi)\text{d}\pi$ instead of $E\left(\int_{\pi\in\Pi}\pi\tau(\pi)\text{d}\pi\right)$. 
In fact, we have the following result:
\begin{proposition}
    If $\tau(\pi)$ is a distribution over $\Pi$, and $E$ is defined as exploitability, then we have
    $$\int_{\pi\in\Pi}\tau(\pi)E(\pi)\text{d}\pi\geq E\left(\int_{\pi\in\Pi}\pi\tau(\pi)\text{d}\pi\right).$$
    \label{prop1}
\end{proposition}
\vspace{-10pt}
The proof is provided in Appendix \ref{app:proof1}. Given the proposition above, there will be an underestimation if we use this method. Also, using maximum alone abandons almost all the information of nearby trajectories, which makes the approximation unstable.
To resolve these problems, we use mean instead of maximum. 
Here, we restate the definition of the exploited level (EL) as
$$EL(\tau)=\mathbb{E}_{\pi}\left[-r(\pi,\pi(\tau))\mid r(\pi,\pi(\tau))\leq 0\right].$$
Except for the conditions mentioned above, the algorithm is mainly based on the following assumption: 
$$E(\tau)\propto EL(\tau)= \frac{\int_{\pi\in\Pi}(-r(\pi,\pi(\tau))^{+}\text{d}\pi}{\int_{\pi\in\Pi}\mathds{1}_{r(\pi,\pi(\tau))\leq 0}\text{d}\pi},$$
where $r(\pi,\pi')$ returns the reward of a player with strategy $\pi'$ by default, $(x)^+=\max{\{x,0\}}$ and $\mathds{1}_{c}=1$ if and only if condition c is satisfied, otherwise $\mathds{1}_{c}=0$. 
The above function means that given a trajectory $\tau$, the mean negative reward of the trajectories with a representation near $\tau$ and reward less than $0$ is proportional to exploitability. 
The right-hand side value is a reasonable measure of a trajectory, which is shown in the toy model. To estimate EL with latent representation space, we provide an alternative definition of $EL_\delta$: 
$$EL_\delta(\tau)=\frac{\sum_{d(f(\tau),f(\tau'))<\delta}(-r(\hat{\pi}, \pi(\tau')))^+}{\sum_{d(f(\tau),f(\tau'))<\delta}\mathds{1}_{r(\hat{\pi}, \pi(\tau'))\leq 0}}.$$
It is obvious that $\lim_{\delta\to 0^{+}}EL_\delta(\tau)=EL(\tau)$.
The property of EL satisfies our requirement that the trajectories that perform similarly to Nash Equilibrium can be detected with an EL near $0$ since we have the following proposition.
\begin{proposition}\label{prop2}
    Given a trajectory $\tau$ and its corresponding distribution $\tau(\pi)$ over $\Pi$, $\pi(\tau)$ is $\epsilon_1$-Nash equilibrium, and we assume that any pure strategy can exploit another strategy by at most $M$. 
    By the smoothness of $f$, we also assume that if $d(f(\tau_1),f(\tau_2))<\delta$, then $\int_{\pi\in\Pi}|\tau_1(\pi)-\tau_2(\pi)|\text{d}\pi<\alpha\delta$, where $\alpha$ is a constant. We have the following result:
    $$EL_\delta(\tau)<\epsilon_1+\alpha\delta M.$$ 
\end{proposition}

Since EL is the average of values satisfying conditions with distance constraints on the representation space, we can train an operator $L$ to estimate EL from representation. We have representation $l$ and reward $r$ for each trajectory $\tau$, and we intend to minimize $\sum_{r^i\geq 0}||L(l^i)-r^i||_2$ so that the prediction from $L(l)$ becomes close to the mean of satisfying reward $r\geq 0$ nearby. We use a two-layer MLP as $L$. 
After training $L$, we can directly obtain EL of a single trajectory $\tau$ even without the reward information. 
By applying representation estimator $f$ and EL estimator $L$ to the trajectory $\tau$, we can get the desired result $L(f(\tau))$.

\subsection{EL Augmentation for Offline Learning}
As described in Section \ref{subsec:ele}, an EL value approaching 0 indicates the trajectory is approaching the Nash equilibrium behavior.
Consequently, we formulate Exploited Level Augmentation (ELA) for the offline learning objective as follows, emphasizing trajectories with a small EL:
$$\mathcal{L}^{\text{ELA}}(\pi)=\mathbb{E}_{\tau}\left[\mathds{1}(EL(\tau)< EL_{\text{thresh}})\cdot\mathcal{L^{\text{OL}}}(\pi, \tau)\right],$$
where $EL_{\text{thresh}}$ is a threshold that specifies the minimum value of an $EL$ suitable for training. 
It provides data by sampling only for trajectories smaller than this value. $\mathcal{L}^{\text{OL}}$ represents an arbitrary method that allows offline learning by leveraging a trajectory such as imitation learning or offline RL methods. 
For example, when incorporating ELA with behavior cloning, the objective function is formulated as:
$$\mathcal{L}^{\text{ELA}}(\pi)=\mathbb{E}_{\tau}\left[\mathds{1}(EL(\tau)< EL_{\text{thresh}})\cdot\sum_{t=0}^{|\tau|}\log\pi(a_t\mid o_t)\right].$$
Note that when employing a maximum value of $EL$ in the dataset as a threshold, it reduces to the original offline learning algorithm.

\begin{figure*}[t]
  \centering
  \subfloat[Player strategy]
  {
    \label{fig:rps1}\includegraphics[width=0.24\textwidth]{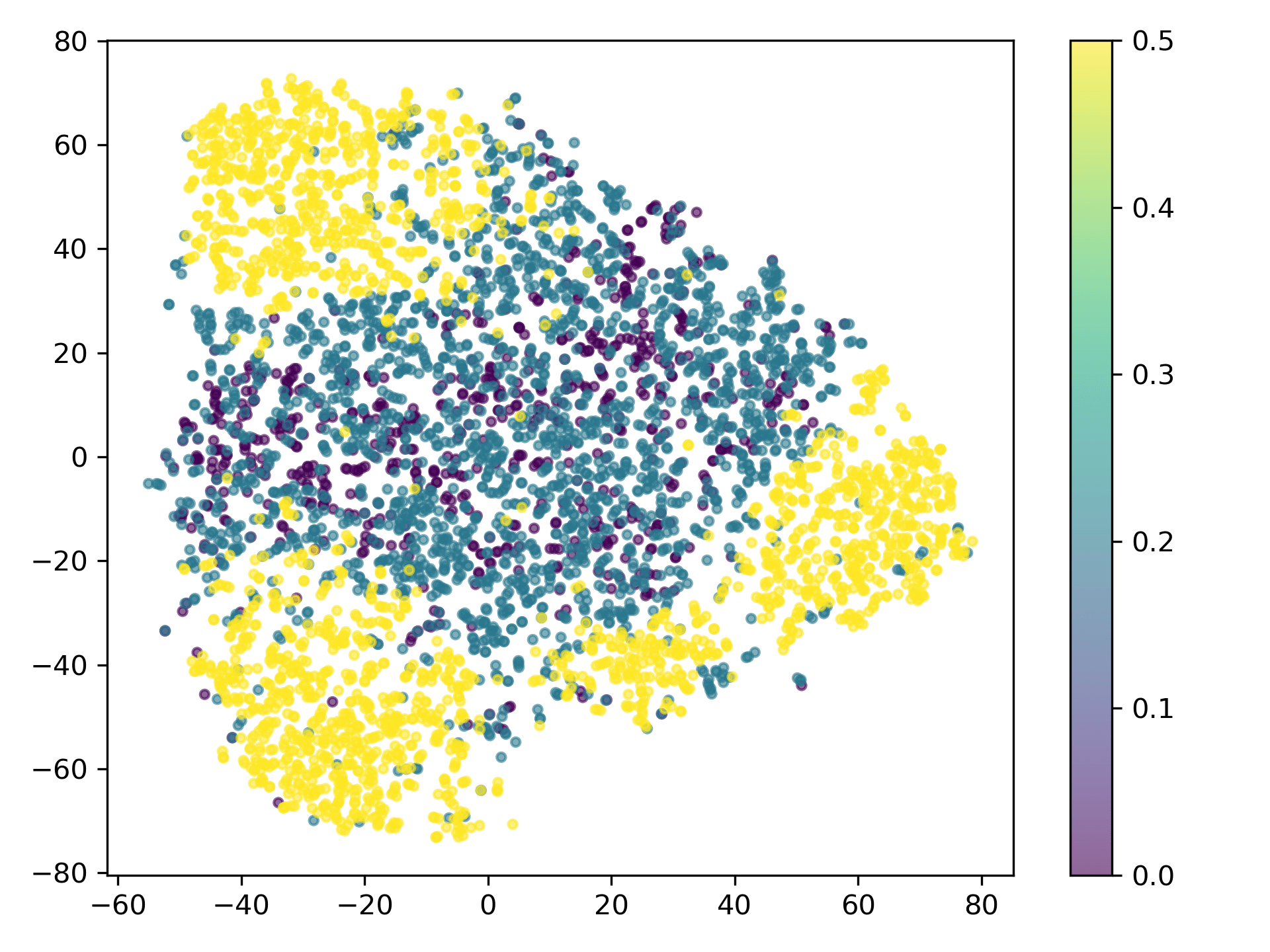}
  }
  \subfloat[Exploited level]
  {
    \label{fig:rps2}\includegraphics[width=0.24\textwidth]{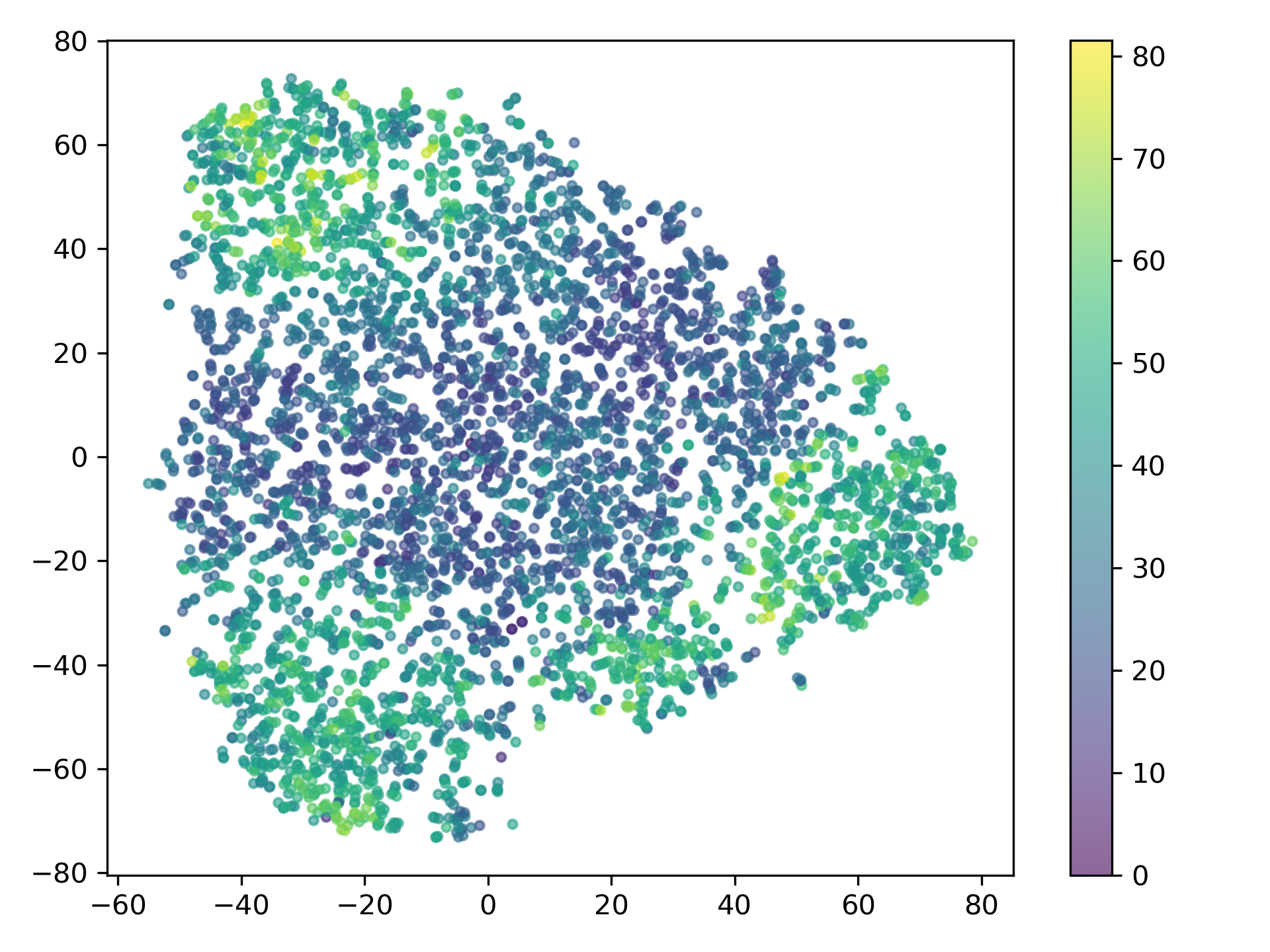}
  }
  \subfloat[Trajectory reward]
  {
    \label{fig:rps3}\includegraphics[width=0.24\textwidth]{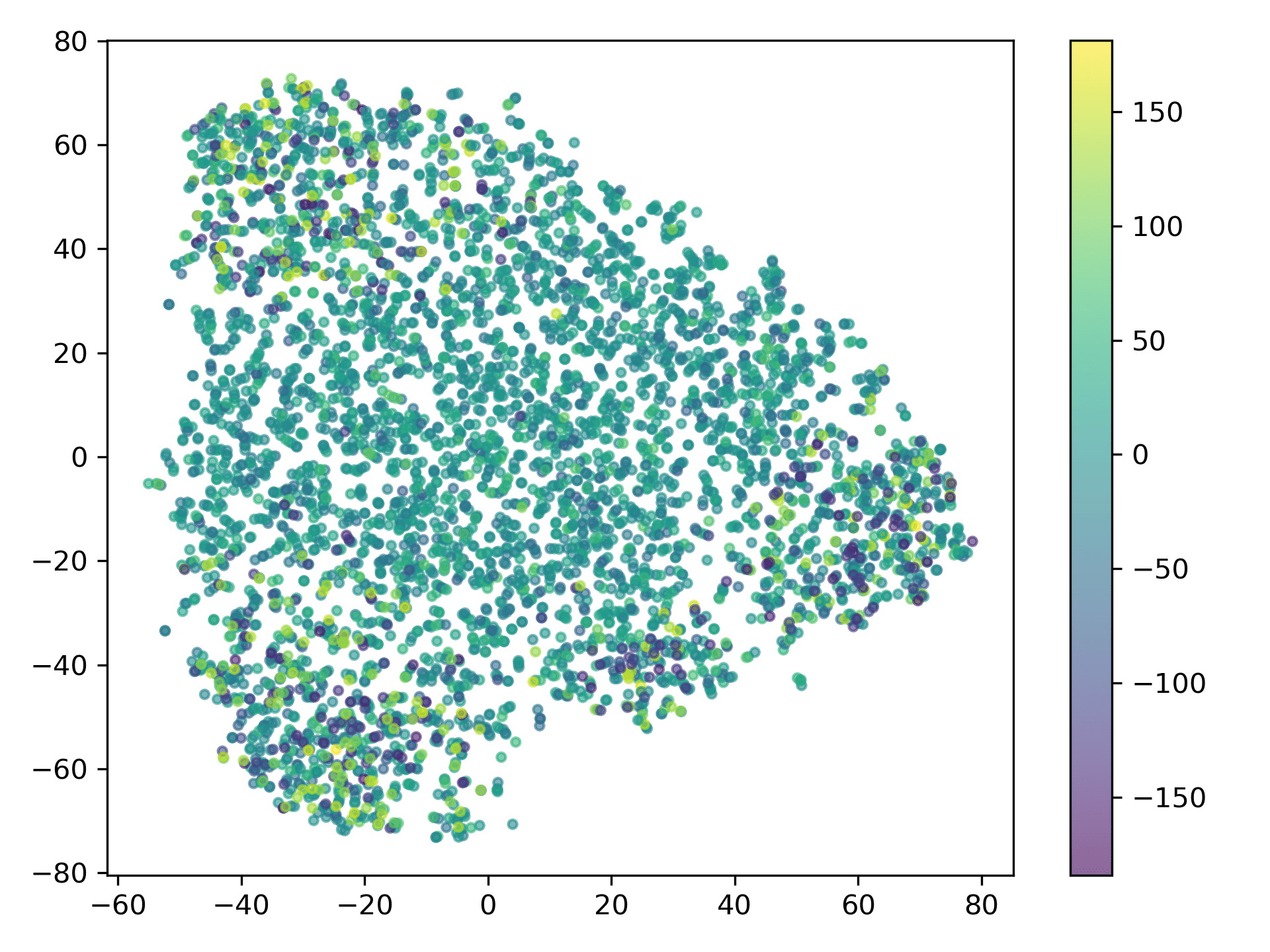}
  }
  \subfloat[3D visualization]
  {
    \label{fig:3d}\includegraphics[width=0.24\textwidth]{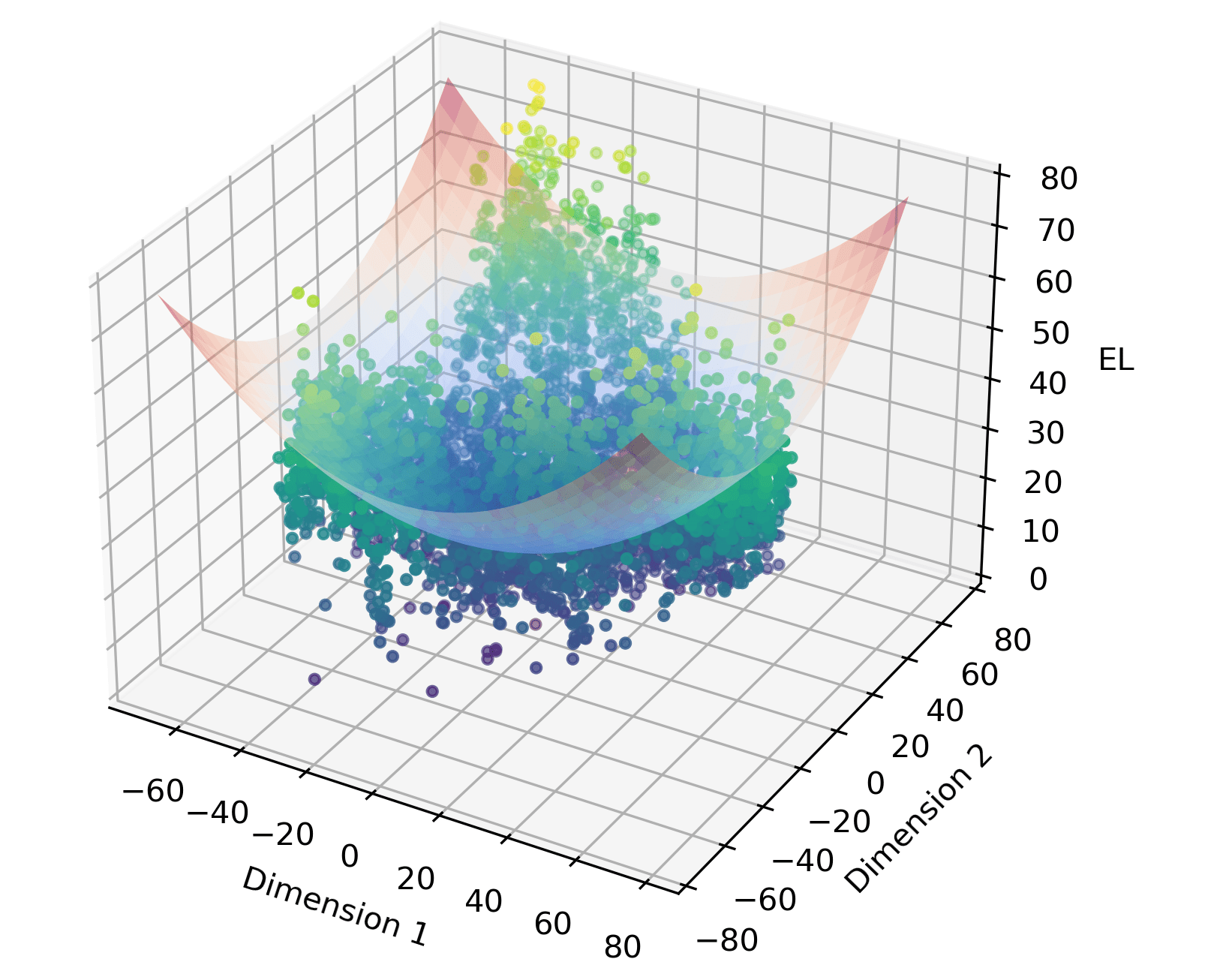}
  }\\
  \vspace{-0.15cm}
  \subfloat[Exploited level]
  {
    \label{fig:pong2}\includegraphics[width=0.24\textwidth]{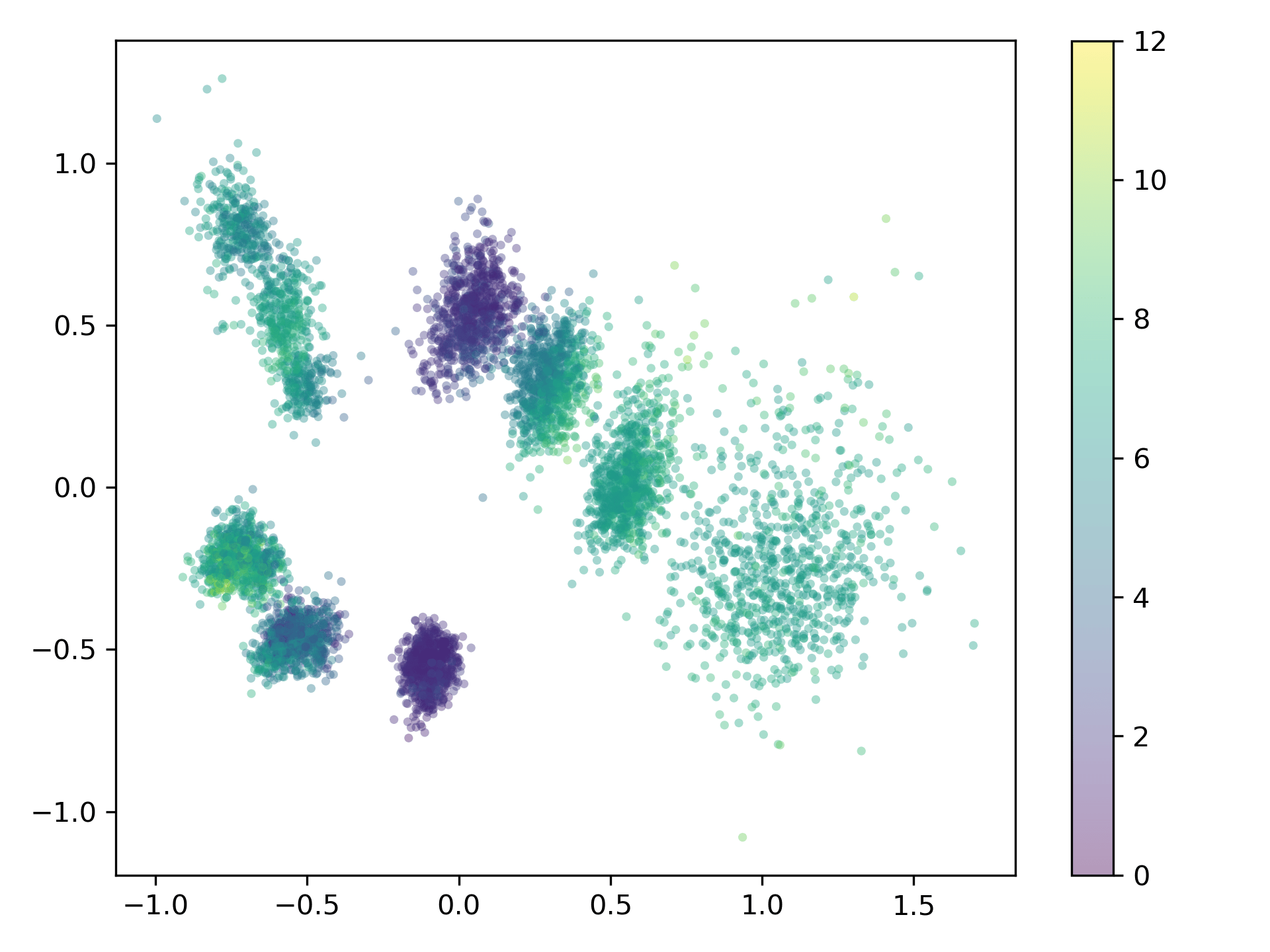}
  }
  \subfloat[Trajectory reward]
  {
    \label{fig:pong3}\includegraphics[width=0.24\textwidth]{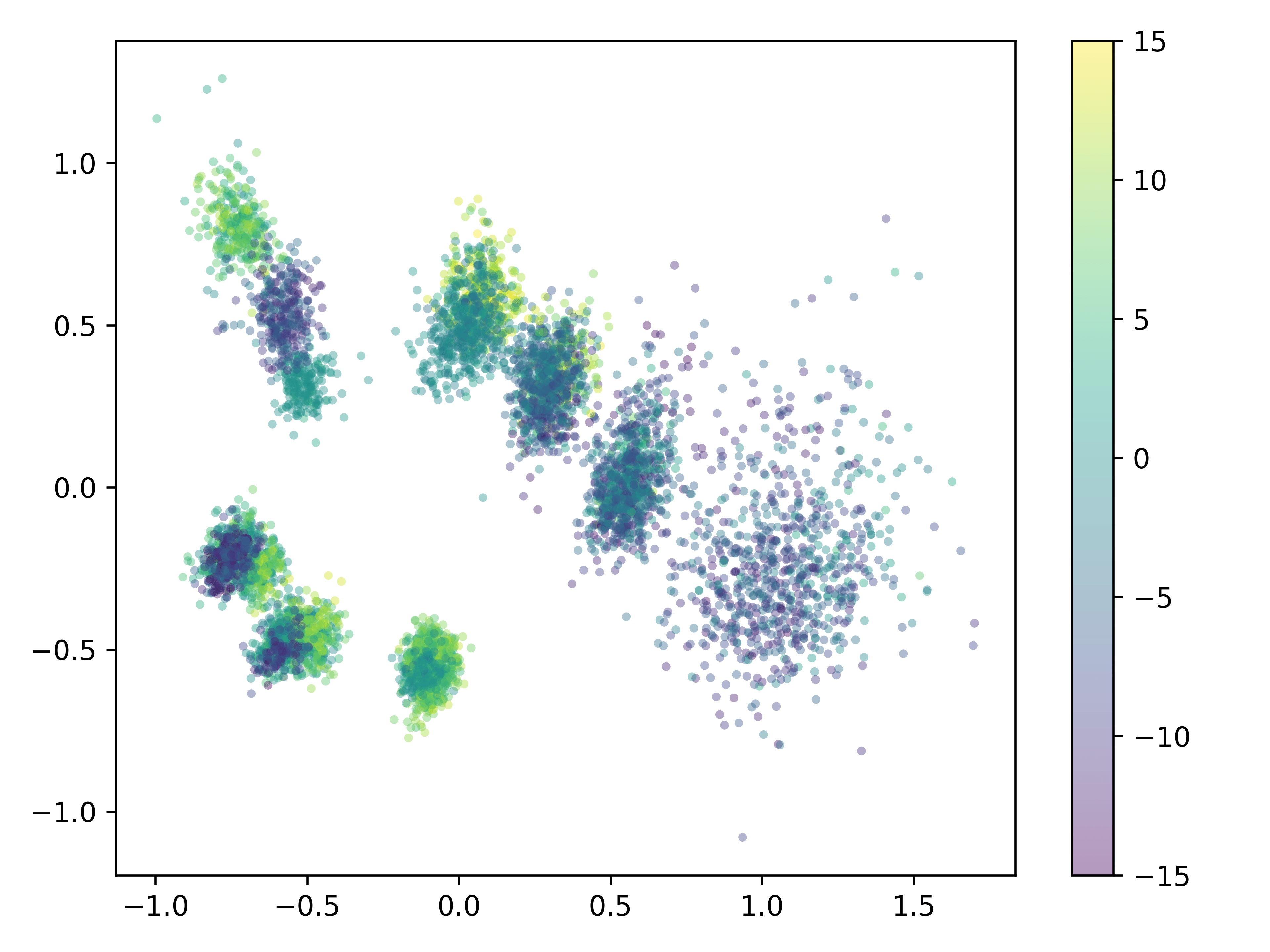}
  }
  \subfloat[Exploited level]
  {
    \label{fig:poker2}\includegraphics[width=0.24\textwidth]{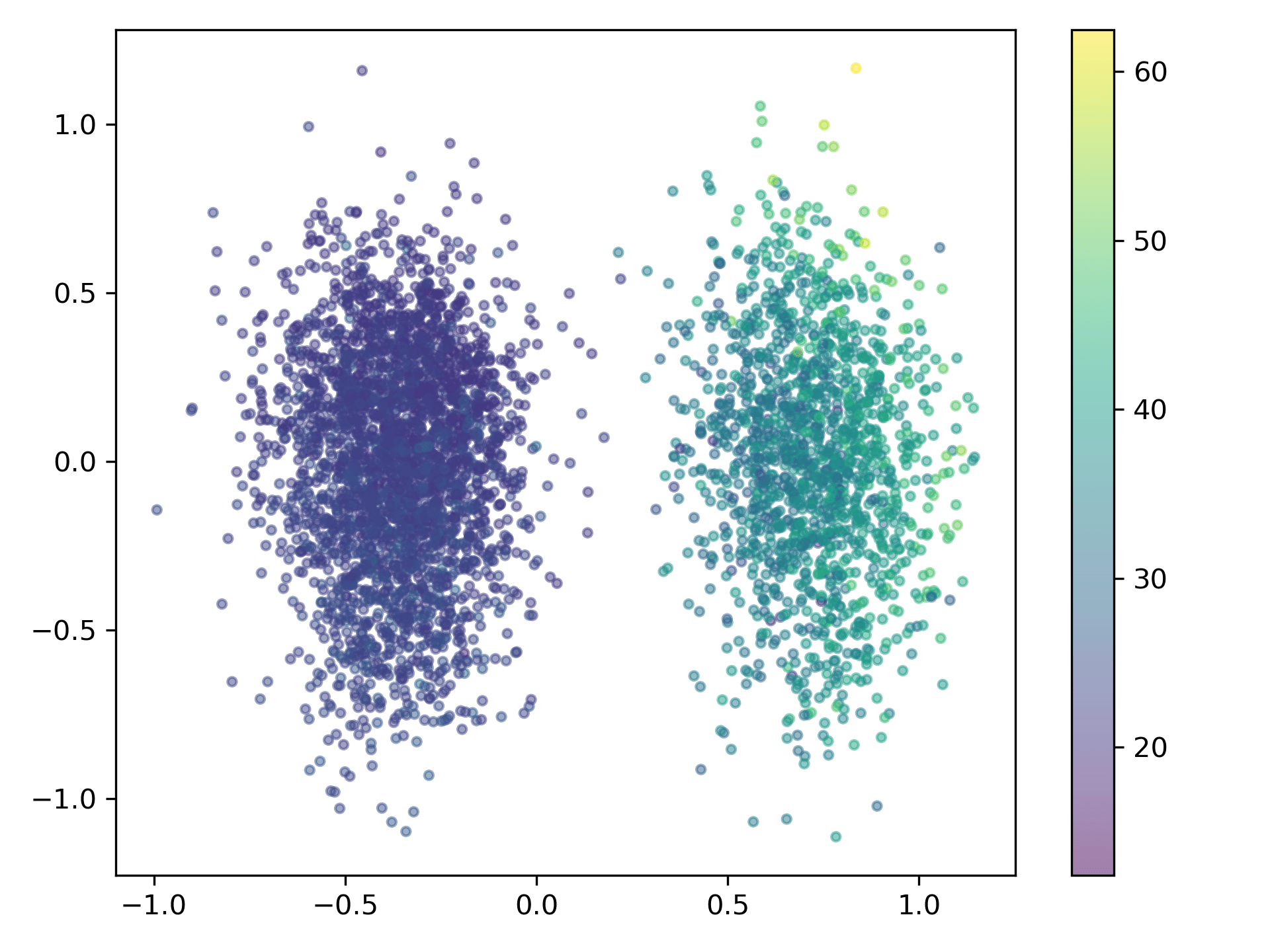}
  }
  \subfloat[Trajectory reward]
  {
    \label{fig:poker3}\includegraphics[width=0.24\textwidth]{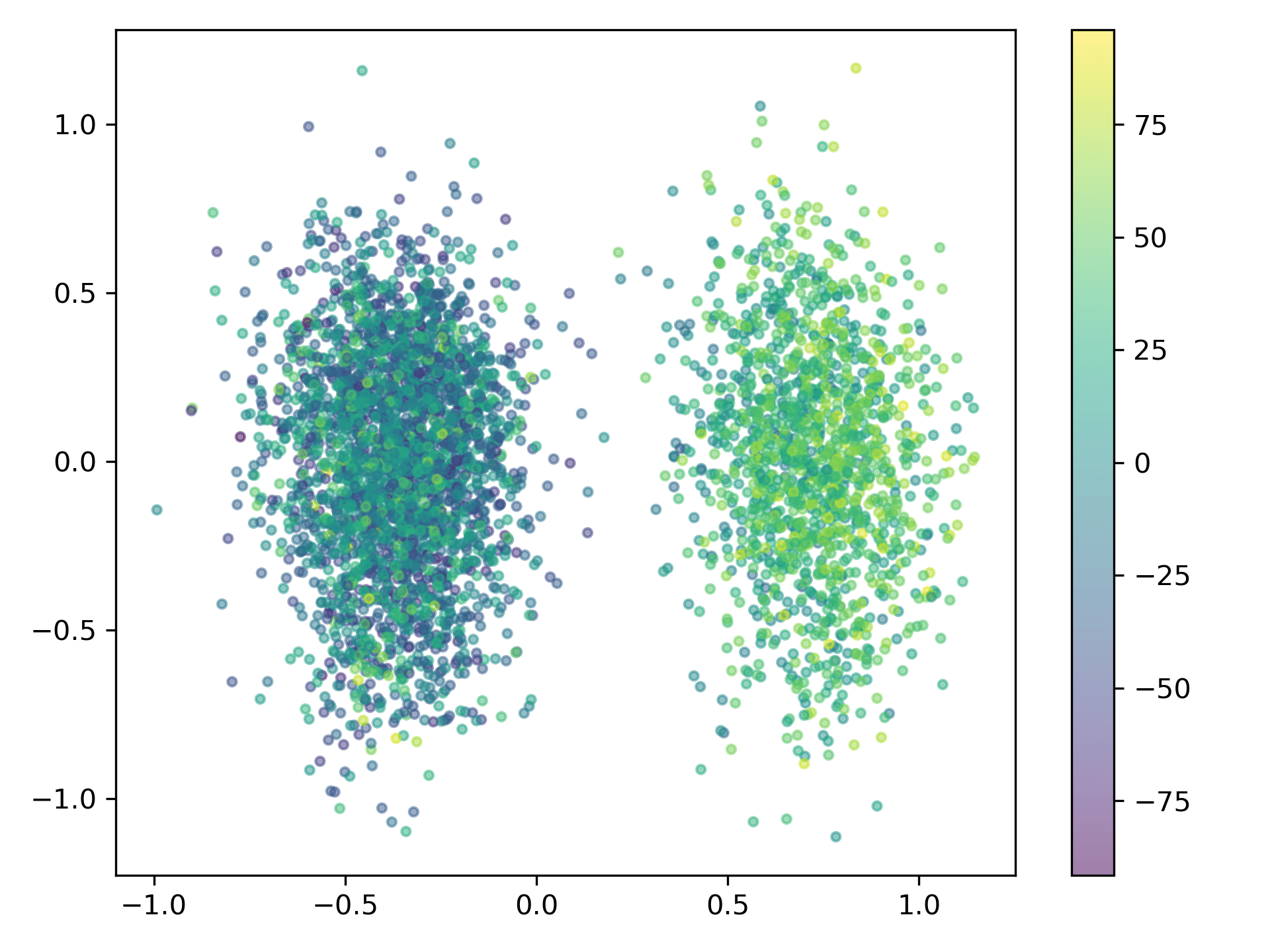}
  }
  \vspace{-0.05cm}
  \caption{The trajectory representations of the \textbf{(a-d)} RPS, \textbf{(e-f)} Two-player Pong, and \textbf{(g-h)} Limit Texas Hold’em.}
  \vspace{-0.2cm}
  \label{fig:rps}
\end{figure*}

\section{Experiments}
\subsection{Experiment Settings}
We use two-player zero-sum games to validate the effectiveness of our approach: Rock-Paper-Scissors (RPS), Two-player Pong, and Limit Texas Hold’em \cite{zha2020rlcard}, which are introduced in Appendix \ref{app:games}. The implementation details for our method are provided in Appendix \ref{app:details}.

\textbf{Dataset generation.} We employ different methods to create training datasets with diverse demonstrators for the environments.
For RPS, we choose the strategy to generate trajectories for RPS as a random strategy with a preference for action $a$ with bias $p$, where $\pi(a')=(1-p)/3$ and $\pi(a)=(1+2p)/3$ $\forall a'\neq a$.
For Pong, we use self-play with opponent sampling \cite{bansal2018emergent} with the Proximal Policy Optimization (PPO) algorithm \cite{schulman2017proximal}.
For Limit Texas Hold'em, we use neural fictitious self-play \cite{heinrich2016deep} with Deep Q-network (DQN) algorithm \cite{mnih2013playing} to generate expert policies, given its complexity and the need to adapt to various opponents. 
Behavior models are then selected from multiple intermediate checkpoints to generate the offline data.

\textbf{Evaluation metrics.}
We evaluated our method across three environments to estimate the representation and EL of trajectories.
Specifically, we conducted tests in a Rock-Paper-Scissors (RPS) environment to provide insight into the higher-level geometrical representation of strategies, aligning with the hypothesis in \cite{czarnecki2020real}.
In evaluating ELA for offline learning algorithms, we compared average scores over 500 games between two players. For the Two-player Pong, the average score was calculated using the formula $(N_{\text{win}} - N_{\text{lose}}) / N_{\text{game}}$. For Limit Texas Hold’em, since a player can win with different margins depending on a certain game, 
we determined the average score as the difference between the total chips won and lost divided by the total number of games played.

\subsection{Representation and EL Estimation}

The strategy representation of trajectories is firstly reduced to two dimensions using t-SNE (PCA is used in Two-player Pong to preserve scale for our further analysis), followed by coloring based on different labels.

\textbf{RPS.} We demonstrate the results in the RPS in Figure \ref{fig:rps}. 
In RPS, a more biased strategy deviates further from the Nash equilibrium, leading to worse performance.
In Figure \ref{fig:rps1}, we color the representations by the player strategy of each trajectory: the trajectories with bias $0.5$, $0.2$, and $0$ are colored with chartreuse, cyan, and purple, respectively.
We use consistent colors in other images in Figure \ref{fig:rps} to indicate the label value of each plot. 
Figure \ref{fig:rps2} shows the estimated EL derived from P-VRNN and the EL estimator. Figure \ref{fig:rps3} is labeled by the terminal reward of each trajectory.
From the three figures, it is evident that our EL estimator shows similar pattern with Figure \ref{fig:rps1}, considered a ground truth. Furthermore, a decrease in the estimated EL is observed as the trajectory approaches the Nash policy.

Considering the structure of the strategy space, we can analyze an $(N+1)$-dimensional space with $N$ dimension of trajectory representation and one dimension of EL. 
To simplify the problem and make it perceptible, we consider reducing the dimension to $2+1$, as shown in Figure \ref{fig:3d}. Despite not matching precisely, we can see a spinning top structure as predicted. There are more strategies with higher EL that are non-transitive.

\textbf{Two-player Pong and Limit Texas Hold’em.} We also illustrate the results of the Two-player Pong and the Limit Texas Hold'em. 
In the Two-player Pong, we choose eight players with strategies trained by PPO with different checkpoints. 
As shown in Figure \ref{fig:pong2} and \ref{fig:pong3}, the strategy representation is naturally separated into eight clusters, revealing both EL and reward distributions.
Figure \ref{fig:pong2} demonstrates that EL better reflects the strength of each player, as the values within each cluster are more consistent. Additionally, it is observed that the most expansive cluster with the lowest density has the highest EL, suggesting that the least trained strategy exhibits unstable behavior.
In the Limit Texas Hold'em, there are three players—two experts with slightly different strategies and one relatively novice player.
As shown in Figure \ref{fig:poker2} and \ref{fig:poker3}, if we use EL as a filter, most of the trajectories played by experts are retained. However, in Figure \ref{fig:poker3}, there are a lot of greenish points in the cluster on the left, which has a low reward. Therefore, if we filter the reward with a neutral value near $0$, a lot of expert trajectories would be erroneously excluded. 
Figure \ref{fig:poker3} indicates that there is a possibility of expert players obtaining low rewards, which disrupts filtering with reward, while Figure \ref{fig:poker2} demonstrates that the EL estimator can overcome this challenge by changing the reward filter to EL filter.

\begin{figure}[t]
  \centering
  \includegraphics[width=\columnwidth]{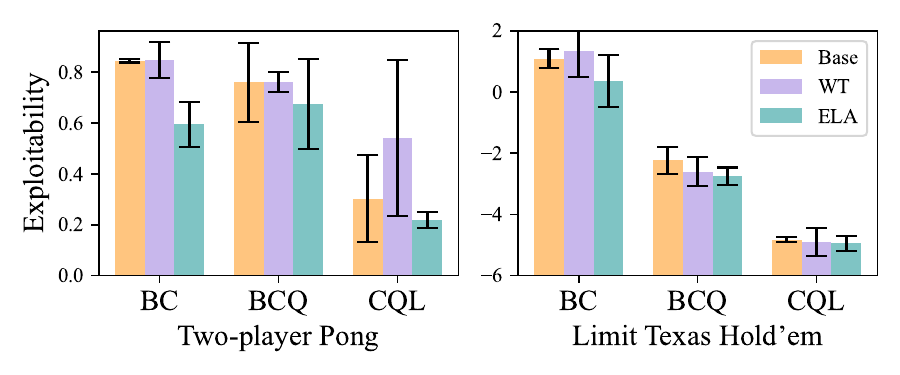}
  \vspace{-0.9cm}
  \caption{Exploitability supported on the demonstrator set generating the offline dataset. Lower is better.}
  \label{fig:exploitability}
\end{figure}

\subsection{EL Augmented Offline Learning}
In our evaluation of the EL-augmented offline learning approach, we considered two main categories of methodologies. For IL, we employed BC, while in the domain of offline RL, we adopted representative algorithms BCQ \cite{fujimoto2019off} and CQL \cite{kumar2020conservative}.
In our evaluation, we excluded methods that rely on online interactions (e.g., GAIL) or necessitate interactions with experts (e.g., DAgger) in offline learning approaches.

\begin{figure}[t]

  \captionsetup[subfloat]{captionskip=-0.1cm, farskip=-0.05cm}
  \centering
  \subfloat[Two-player Pong]
  {
  \label{fig:cross_eval_pong}\includegraphics[width=\columnwidth]{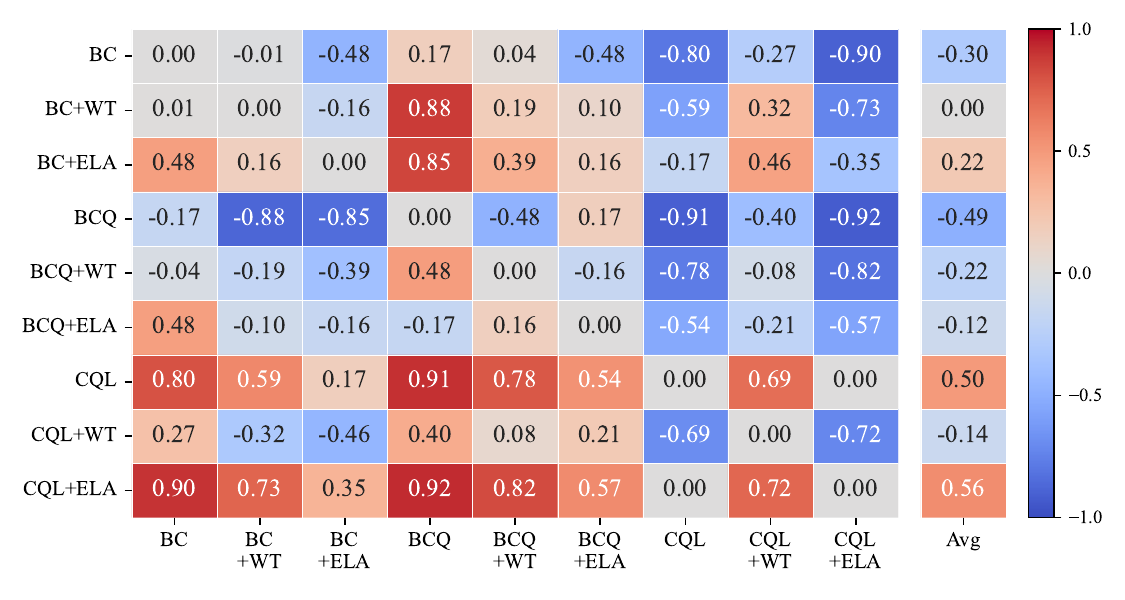}
  }\\
  \subfloat[Limit Texas Hold’em]
  {
    \label{fig:cross_eval_card}\includegraphics[width=\columnwidth]{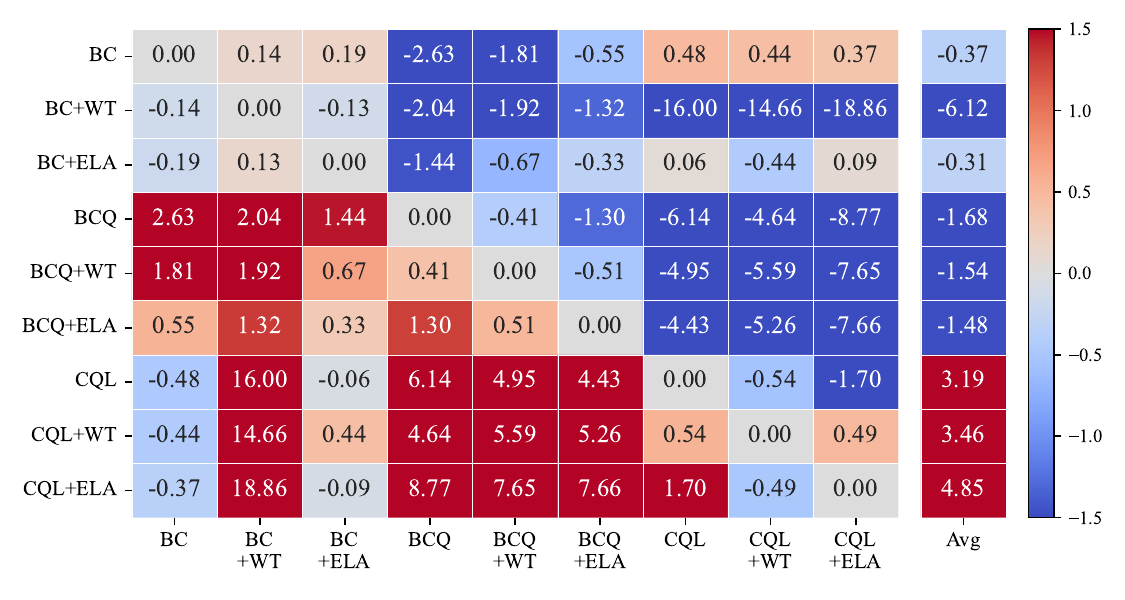}
  }
  \vspace{-0.1cm}
  \caption{Cross evaluation of offline learning algorithms in zero-sum games. Higher is better.}
  \label{fig:cross_Eval}
\end{figure}

We applied ELA to each algorithm to evaluate its performance enhancement. We scaled the learned estimated EL from 0 to 1 by using the maximum and minimum EL in the dataset. A hyperparameter search was conducted to identify the appropriate $EL_\text{thresh}$ for each model and environment. As an additional baseline for ELA, we trained the offline learning algorithm by exclusively selecting the winning trajectory (WT) from the dataset.
In Figure \ref{fig:exploitability}, a comparison of exploitability is presented, supported on the demonstrator set outlined in Section \ref{sec:preliminaries}. Basically, offline RL algorithms show better performance than the imitation learning approach on average because of the offline datasets from mixed demonstrators. Notably, ELA consistently outperforms alternative methods. While WT enhances the performance of the original offline algorithm in some cases, it occasionally hinders performance due to Q-value overestimation stemming from data bias by only selecting the winning trajectory.
Furthermore, to illustrate the relative performance among the trained models, Figure \ref{fig:cross_Eval} displays the outcomes of cross-evaluating various algorithm combinations in the two environments. The value of each cell signifies the score of the model along the horizontal axis compared to the model along the vertical axis. The last column in each subfigure highlights that ELA enhances the performance of all offline learning algorithms in both environments.

\section{Conclusions}
In this work, we proposed an effective framework, ELA, to enhance offline learning methods in zero-sum games. We designed a P-VRNN network, which shows extraordinary results in identifying the strategy distribution of the trajectories. 
We defined the exploited level for the trajectory to measure proximity to Nash equilibrium and subsequently proposed ELA as a universal method for improving the performance of offline learning algorithms.
We built a solid theoretical foundation for ELA, and the experiments on multiple environments and algorithms showed positive results in adding ELA. We explored a broad variety of algorithms with different hyperparameters.
In future work, we aim to explore zero-sum games with a larger number of players.
We consider utilizing the strategy representation in different ways over a larger variety of environments since the P-VRNN does not require the game to be zero-sum.

\section*{Social Impacts}
Our paper introduces a novel approach to enhance the existing offline learning algorithms in zero-sum games. The improved efficiency in identifying dominant strategies may inadvertently amplify strategic advantages in competitive domains, posing risks to fairness. Ethical considerations are necessary to responsibly deploy the method and mitigate potential negative results in real-world applications.

\bibliography{ref}
\bibliographystyle{icml2024}

\newpage
\appendix
\onecolumn
\section{The Proof of Proposition \ref{prop1}}\label{app:proof1}

\begin{proof}
    For simplicity, we only prove in a 2-player setting. By definition of exploitability, $E(\pi)=-r(BR(\pi),\pi)$. So we have
    \begin{align*}
        E(\pi(\tau)) &= -r\left(BR(\pi(\tau)),\pi(\tau)\right) \\
        &= -r\left(\text{argmax}_{\pi_{-i}}r(\pi_{-i},\pi(\tau)), \pi(\tau)\right) \\
        &= -\int_{\pi\in\Pi}\tau(\pi)r\left(\text{argmax}_{\pi_{-i}}r(\pi_{-i},\pi(\tau)),\pi\right)\text{d}\pi \\
        &\leq -\int_{\pi\in\Pi}\tau(\pi)r\left(\text{argmax}_{\pi_{-i}}r(\pi_{-i},\pi),\pi\right)\text{d}\pi \\
        &= -\int_{\pi\in\Pi}\tau(\pi)r\left(BR(\pi),\pi\right)\text{d}\pi \\
        &= \int_{\pi\in\Pi}\tau(\pi)E(\pi)\text{d}\pi
    \end{align*}
    The inequality is established by the property of $\text{argmax}$ function.
\end{proof}

\section{The Proof of Proposition \ref{prop2}}\label{app:proof2}

\begin{proof}
    Since $\pi(\tau)$ is $\epsilon_1$-Nash equilibrium, the exploitability $E(\pi(\tau))\leq \epsilon_1$. Thus for an arbitrary $\hat{\pi}$, we have $r(\hat{\pi},\pi(\tau))\geq -\epsilon_1$. Hence, for all $\tau'$ satisfying $d(f(\tau),f(\tau'))<\delta$, we have
    \begin{align*}
        r(\hat{\pi},\pi(\tau')) &= \int_{\pi\in\Pi}\tau'(\pi)r(\hat{\pi},\pi)\text{d}\pi \\
        &= \int_{\pi\in\Pi}\tau(\pi)r(\hat{\pi},\pi)\text{d}\pi + \int_{\pi\in\Pi}(\tau'(\pi)-\tau(\pi))r(\hat{\pi},\pi)\text{d}\pi \\
        &\geq r(\hat{\pi},\pi(\tau))-\int_{\pi\in\Pi}\left|\tau'(\pi)-\tau(\pi)\right|\left|r(\hat{\pi},\pi)\right|\text{d}\pi \\
        &\geq -\epsilon_1-M\int_{\pi\in\Pi}\left|\tau'(\pi)-\tau(\pi)\right|\text{d}\pi \\
        &> -\epsilon_1-\alpha\delta M
    \end{align*}
    Thus, we have
    \begin{align*}
        EL_\delta(\tau) &= \frac{\sum_{d(f(\tau),f(\tau'))<\delta}(-r(\hat{\pi}, \pi(\tau')))^+}{\sum_{d(f(\tau),f(\tau'))<\delta}\mathds{1}_{r(\hat{\pi}, \pi(\tau'))\leq 0}} \\
        &\leq \max_{d(f(\tau),f(\tau'))<\delta}-r(\hat{\pi}, \pi(\tau')) \\
        &< \epsilon_1+\alpha\delta M
    \end{align*}
\end{proof}

\section{The Games and Implementation Details}
\subsection{Overview of the Zero-Sum Games}\label{app:games}
We choose the following well-known games in our experiments: 
\begin{itemize}
    \item \textbf{Rock-Paper-Scissors (RPS)}: Players have three potential actions to take: rock, paper, and scissors. The observation of each player is the action of the opponent in the last round. In each trajectory, RPS games are played for $T=500$ times consecutively. The player who wins gets $+1$ point, and the player who loses gets $-1$ point. When there is a draw, the point is not changed.
    \item \textbf{Two-player Pong}: Each player controls a paddle on one side of the screen. The goal is to keep the ball in play by moving the paddles up or down to hit it. If a player misses hitting the ball with their paddle, it loses the game. The observation of players includes ball and paddle positions across two consecutive time steps and potential actions include moving up or down.
    \item \textbf{Limit Texas Hold'em}: Players start with two private hole cards, and five community cards are revealed in each stage (the flop, turn, and river). Each player has to create the best five-card hand using a combination of their hole and the community cards. During the four rounds, players can select call, check, raise, or fold. The players aim to win the game by accumulating chips through strategic betting and building strong poker hands. The observation of players is a 72-element vector, with the first 52 elements representing cards (hole cards and community cards) and the last 20 elements tracking the betting history in four rounds.
\end{itemize}
\subsection{Implementation Details}\label{app:details}
In the actual implementation of P-VRNN, the action $a_t$ and observation $o_t$ pass through neural networks $\psi_\text{a}$ and $\psi_\text{o}$ first to reduce dimension and extract features.
The functions $\phi_{\text{p}}$, $\phi_{\text{e}}$, and $\phi_{\text{d}}$ are implemented with multi-layer perceptron (MLP) with latent space dimension $z_\text{dim}=8$, hidden layer dimension $h_\text{dim}=8$, recurrence layer dimension $r_\text{dim}=8$ and representation dimension $l_\text{dim}=8$. Gated Recurrent Unit (GRU) \cite{chung2014empirical} is used as the recurrence function $\phi_{\text{r}}$. We trained the models for $100$ epochs with a learning rate of $0.0005$ and a batch size of $32$ trajectories using the Adam optimizer.
As for EL estimation, we also use GRU, and we use the recurrent output of the final step as function output. 

In our offline learning experiments, we utilize an MLP architecture for the actor network, with two hidden layers of $256$ units each. Our offline dataset consists of $45\text{K}$ trajectories, each containing $100$ time steps for the Limited Texas Hold'em game and $1\text{K}$ time steps for the Two-player Pong game. During offline learning, we trained the models for $300$ epochs with a learning rate of $0.0005$. We set the minibatch number to $50$ for each epoch, employing the Adam optimizer to ensure a consistent number of updates for all methods. We used a widely used codebase for BCQ\footnote{https://github.com/sfujim/BCQ} and CQL\footnote{https://github.com/BY571/CQL} to ensure consistency and reproducibility.
All experiments were conducted using an RTX 2080 Ti GPU and an AMD Ryzen Threadripper 3970X CPU.

\section{Remarks on Learning Strategy Representation of Trajectories in Multi-Agent Games}\label{app:rmk}

In an imperfect information game, relying on models that only consider the observation and action information of a single time step is insufficient to obtain strategy representations for trajectories.
Also, directly inferring representation from observation and action using an operator introduces bias due to the influence of opponents' strategies on their actions. Unlike a single player in a specific environment, where actions directly affect observations, the presence of multiple players leads to diverse observations even when the agent's behavior remains constant.
For the same reason, the representation should remain independent of the reward, as it is a combined outcome of both players, including the opponent. Furthermore, the representation should also be informative enough to predict a player's subsequent action.
Therefore, we turn to variational recurrent neural networks (VRNN) \cite{NIPS2015_b618c321}, which is widely used for sequential generation to enable such prediction.

\end{document}